\documentclass[10pt,conference]{IEEEtran}
\usepackage{graphicx}

\begin{document}


\title{Quantum key distribution based on a Sagnac loop interferometer and polarization-insensitive phase modulators}




\author{\authorblockN{Bing Qi, Lei-Lei Huang, Hoi-Kwong Lo, Li Qian}
\authorblockA{Center for Quantum Information and Quantum Control\\
Dep. of Physics and Dep. of Electrical and Computer Engineering\\
University of Toronto, Toronto, Canada M5S 3G4\\
Email: bqi@physics.utoronto.ca, leilei.huang@utoronto.ca\\
hklo@comm.utoronto.ca, l.qian@utoronto.ca}}


\maketitle

\begin{abstract}

We present a design for a quantum key distribution (QKD) system in a
Sagnac loop configuration, employing a novel phase modulation scheme
based on frequency shift, and demonstrate stable BB84 QKD operation
with high interference visibility and low quantum bit error rate
(QBER). The phase modulation is achieved by sending two light pulses
with a fixed time delay (or a fixed optical path delay) through a
frequency shift element and by modulating the amount of frequency
shift. The relative phase between two light pulses upon leaving the
frequency-shift element is determined by both the time delay (or the
optical path delay) and the frequency shift, and can therefore be
controlled by varying the amount of frequency shift. To demonstrate
its operation, we used an acousto-optic modulator (AOM) as the
frequency-shift element, and vary the driving frequency of the AOM
to encode phase information. The interference visibility for a 40km
and a 10km fiber loop is 96\% and 99\%, respectively, at single
photon level. We ran BB84 protocol in a 40-km Sagnac loop setup
continuously for one hour and the measured QBER remained within the
2\%$\sim$5\% range. A further advantage of our scheme is that both
phase and amplitude modulation can be achieved simultaneously by
frequency and amplitude modulation of the AOM's driving signal,
allowing our QKD system the capability of implementing other
protocols, such as the decoy-state QKD and the continuous-variable
QKD. We also briefly discuss a new type of Eavesdropping strategy
(``phase-remapping" attack) in bidirectional QKD system.

\end{abstract}

\section{Introduction}

One important practical application of quantum information is
quantum key distribution (QKD), whose unconditional security is
based on the fundamental law of quantum mechanics [1-6]. In
principle, any eavesdropping attempt by a third party (Eve) will
unavoidably introduce quantum bit errors, so, it's possible for the
legitimate users (Alice and Bob) to upper bound the amount of
information acquired by the eavesdropper from some system parameters
and the measured quantum bit error rate (QBER). Alice and Bob can
then distill out a final secure key by performing error correction
and privacy amplification. Because Alice and Bob can't distinguish
the intrinsic QBER due to imperfections in a practical QKD system
from the one induced by Eve, to guarantee the unconditional
security, they have to assume all errors originate from
eavesdropping. Obviously, the QKD system with higher intrinsic QBER
will yield a lower secure key rate.  As the QBER reaches some
threshold, QKD is not unconditional secure anymore.

In a practical phase-coding QKD system, Alice and Bob achieve phase
encoding/decoding with phase modulators (PM) and Mach Zehnder
interferometers (MZI) [7-8].  However, there are a few practical
difficulties if QKD is to be implemented over long distance through
fiber: namely, phase and polarization instabilities. In this case,
the intrinsic QBER induced by the imperfect interference can be
described as [2]
\begin{equation} \emph{QBER}=(1-\emph{V})/2
\end{equation}
where V is the interference visibility.

Although promising progresses have been achieved by using active
feedback control to stabilize the interferometer [8], the ``plug \&
play" auto-compensating QKD structure employing a Faraday mirror has
demonstrated higher performance in practice [9]. Like the ``plug \&
play" system, the Sagnac loop also offers phase stability and
polarization stability, as the two interfering signals travel
through the same path, but its structure is in principle much
simpler than the ``plug \& play" scheme [10-13]. However, all
reported Sagnac QKD systems employed polarization-sensitive phase
modulators, requiring complicated polarization controls, which makes
this scheme unattractive. For example, four polarization controllers
were employed in [12], and the interference visibility for a 5 km
fiber loop was only 87\%.

In this paper, we present a design for an AOM-based
polarization-insensitive phase modulation scheme together with a
Sagnac QKD system, and demonstrate stable QKD operation over one
hour without feedback control. Although this system is designed for
the BB84 protocol [1], with a few straightforward modifications, it
can also be adapted to implement other protocols, such as the decoy
state QKD [14-20] and the continuous variable QKD [21].

\section{AOM-based phase modulator and bidirectional Sagnac QKD system}

As two light pulses with a fixed time delay pass though a frequency
shift element, a relative phase shift is introduced between the two
pulses, which is determined by the amount of frequency shift. So, by
modulating the frequency shift, phase modulation can be achieved.
This is the basics principle of our phase modulation scheme.

Fig. 1a shows the basic structure of this novel phase modulator,
which consists of an acousto-optic modulator (AOM), followed by a
fiber with length $L$. For the first-order diffracted light, the AOM
will introduce a frequency shift equal to its driving frequency $f$
(due to the Doppler effect). The phase of the diffracted light is
also shifted by an amount of $\phi(t)$ , which is the phase of
acoustic wave at the time of diffraction [22]. Assuming two light
pulses, $S_1$ and $S_2$, are in phase and are sent to the phase
modulator at the same time from opposite directions as shown in Fig.
1a. They will reach the AOM at different times with a time
difference $t_2-t_1=nL/C$. Here $n$ is effective index of fiber and
$C$ is the speed of light in vacuum. The phase difference between
$S_1$ and $S_2$ after they go through the phase modulator will be

\begin{equation}
\Delta \phi=\phi(t_2)-\phi(t_1)=2\pi f(t_2-t_1)=2\pi nLf/C
\end{equation}

\begin{figure}[hbt]
\centering
\includegraphics[width=14cm]{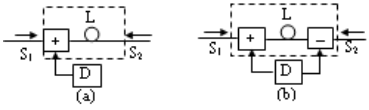}\\
\caption{a) A phase modulator with one frequency up-shifter; b) A
phase modulator with a pair of frequency shifters. +: up-shifting
AOM,  -: down-shifting AOM, L: fiber with length L, D: AOM driver}
\label {modulators}
\end{figure}

By modulating AOM's driving frequency $f$, the relative phase
between $S_1$ and $S_2$ can be modulated. We remark that the
frequency of light will be up-shifted by this phase modulator by an
amount $f$. To remove this ``side-effect", we can add another
frequency down shifter at the other end of the fiber, as shown in
Fig.1b. In Fig.1b, the two AOMs, which are driven by the same
driver, will shift the frequency of light by the same amount but
with different signs. So the net frequency shift will be zero. Since
a down-shift AOM will shift the phase of the diffracted light by
$-\phi(t)$, the resulting phase difference between $S_2$ and $S_1$
after they go through the phase modulator will be

\begin{equation}
\Delta \phi=\phi_{S_2}-\phi_{S_1}=4 \pi nLf/C
\end{equation}

Compared with the $LiNbO_3$ waveguide-based phase modulator, the
AOM-based phase modulator we proposed can be designed to be
insensitive to the polarization state of the input light. This could
dramatically simplify the design of many QKD systems. As shown in
(2) and (3), the phase delay is determined by the acoustic frequency
$f$, which can be controlled very precisely (1ppm frequency
resolution is quite common). This implies that a high resolution
phase modulation can be achieved with our design. As a comparison,
the resolution of $LiNbO_3$ phase modulator is in the order of 1\%.

We have proposed a Sagnac QKD system employing this novel phase
modulator, as shown in Fig.2. To realize the BB84 protocol, Alice
randomly encodes the relative phase between the clockwise and
counterclockwise light pulses with the AOM-based phase modulator
PM1, while Bob randomly chooses his measurement basis with phase
modulator PM2. In Alice's side, two classical photo detectors (PD1,
PD2) and two wavelength filters (F1, F2) are introduced to counter
Trojan horse attack.  These photo detectors could also be used for
synchronization purpose.

We remark the transmittance of AOM can also be modulated by
modulating the amplitude of its driving signal, so, the same device
can function as an amplitude modulator as well as a phase modulator.
(We remark that the newly developed decoy-state QKD protocol
[14-20], which improves the secure key generation rate of practical
QKD system dramatically, can be easily realized in our proposed
setup. In decoy-state QKD, Alice randomly adds decoy pulses, which
are used for testing the communication channel, into the signal
pulses for key distribution. The decoy pulses are identical to the
signal pulses except for the average photon number. In the first
experimental demonstration of the decoy QKD [20], we added an
external AOM into a ``plug \& play" system to randomly modulate the
amplitude of each laser pulse according to a random profile. Because
the phase modulator we proposed can also function as an amplitude
modulator, our proposed QKD system can implement decoy state
protocol easily: Alice can achieve phase modulation by modulating
the frequency of the AOM driving signal, at the same time, she can
also modulate the intensity of each pulse by modulating the
amplitude of the driving signal.)

\begin{figure}[hbt]
\centering
\includegraphics[width=14cm]{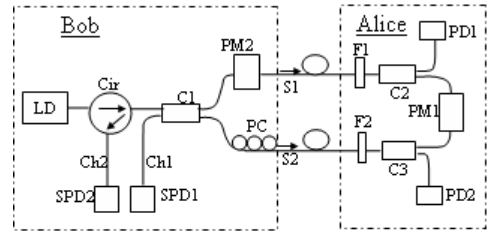}\\
\caption{Proposed QKD system: LD-pulsed laser diode; Cir-circulator;
C-2x2 coupler; PC-polarization controller; F1,F2-optical filter;
PD1, PD2-classical photo detector; PM1, PM2- AOM-based phase
modulator (as shown in Fig.1b); SPD1, SPD2-Single Photon detector}
\label {Proposed}
\end{figure}

We remark, by introducing asymmetrical attenuation in the fiber loop
(for example, fiber isolators) and replacing the SPDs with homodyne
detector, continuous variable QKD [21] could be implemented with the
proposed system: here, Alice can randomly modulates the amplitude
quadrature and phase quadrature with PM1, while Bob chooses which
quadrature to measure with PM2.

\section{Experimental results with a simplified system}

To demonstrate the feasibility of our design, we have performed
experiments with a simplified version of our proposed system shown
in Fig.3. Here, only Alice holds a phase modulator, while Bob always
chooses the same measurement basis. Strictly speaking, no genuine
secret key can be distributed by this simplified system, but it
allows us to evaluate the performance of a fully developed system
(as shown in Fig.2) in terms of stability and quantum bit error rate
(QBER). In Fig.3, the cw output from a \emph{1550nm} laser (L) is
modulated by an amplitude modulator (AM) to generate 500ps laser
pulses. Each laser pulse is split into $S_1$ and $S_2$ at a
symmetric fiber coupler, which go through a long fiber loop
($L1+L2\approx40km$) in the clockwise and counterclockwise
directions, respectively. The interference patterns at Ch1 and Ch2
are measured by two InGaAs single photon detectors (SPD, Id
Quantique, id200), which work in gated mode. For a 5ns gating
window, the overall detection efficiency is $\sim$10\% and the dark
count probability is  $5\times10 ^{-5}$per gating window. A
fiber-pigtailed AOM (Brimrose inc.) is placed inside the fiber loop
asymmetrically ($L1-L2\approx 700m$). Due to this asymmetry, phase
modulation between $S_1$ and $S_2$ can be achieved by modulating
AOM's driving frequency, similar to the phase modulator shown in
Fig.1a. Because of the birefringence in the fiber loop, the
polarization states of $S_1$ and $S_2$ could be different after they
go through the fiber loop [23]. This is compensated by a
polarization controller (PC). The synchronization is achieved as
follows: A pulse generator (PG), which is triggered by a function
generator (FG1), drives the amplitude modulator (AM) to produce
500ps laser pulses. FG1 also triggers a delay generator (DG,
Stanford research system, DG535), which in turn produces two gating
signals for SPD1 and SPD2, and one trigger signal for a data
acquisition card (NI, PCI-6115). The AOM is driven by another
function generator FG2, whose frequency is controlled by the data
acquisition card.

\begin{figure}[hbt]
\centering
\includegraphics[width=14cm]{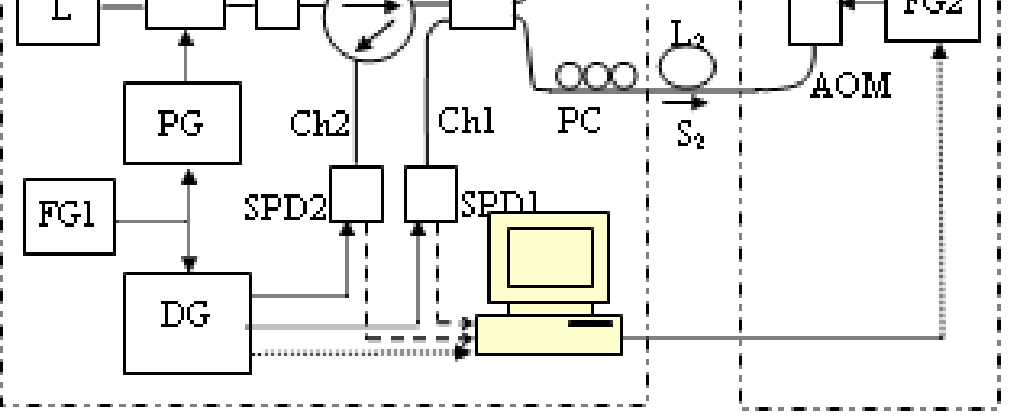}\\
\caption{Experimental setup: L-1550nm cw laser; AM-amplitude
modulator; Att-attenuator; Cir-circulator; C-2x2 coupler;
PC-polarization controller; FG1, FG2-function generator; PG-Pulse
generator; DG-Delay generator; SPD1, SPD2-Single Photon detector}
\label {setup}
\end{figure}

We measured the interference visibility by scanning the frequency of
FG2 while recoding the outputs from the two SPDs. The average photon
number per pulse (out from Alice's side) was set to be 0.8, which
matched with signal photon level in a decoy state QKD system
[19-20].  The measured interference visibility for a 40km and a 10km
fiber loop was 96\% and 99\% respectively.

To run the BB84 protocol, a random number file (1Kbits) is preloaded
to the buffer of the data acquisition card. This random file
contains a sequence of four discrete values corresponding to the
four phase values in the BB84 protocol \{$0$, $\pi/2$, $\pi$,
$3\pi/2$\}. Once triggered, the data acquisition card reads out a
value from the random file and sends it to FG2 to encode Alice's
phase information. The data acquisition card also samples the
outputs from SPDs (Bob's measurement results) into its input buffer.
In this preliminary setup, Bob always uses the same basis for his
measurement. After transmitting 100K bits, Alice and Bob can
estimate the QBER by comparing the data contained in Alice's random
file and Bob's measurement results. We ran the system continuously
for one hour without any adjustment, the QBER drifted slowly from
2\% to 5\%, as shown in Fig.4. In practice, to improve the long-term
stability, simple recalibration process can be employed, as in other
QKD systems. During this experiment, the pulse repetition rate was
set to 1 KHz for easy synchronization.

\begin{figure}[hbt]
\centering
\includegraphics[width=8cm]{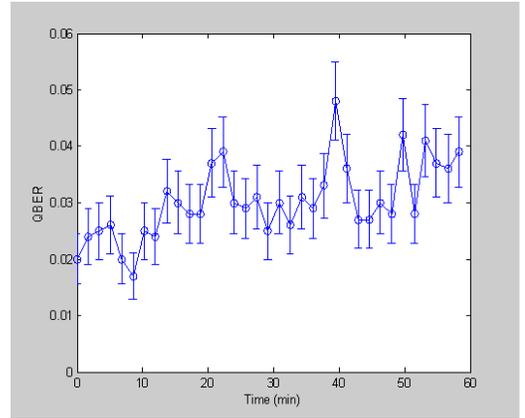}\\
\caption{Measured QBER in one hour without any feedback control. The
photon level is 0.8photon/pulse (The error bars indicate the
statistic fluctuation due to the finite detection events)} \label
{QBER}
\end{figure}

For a typical AOM, its frequency modulation rate is in the range of
1$\sim$10MHz, which is compatible with the operation rate of today's
QKD system. The ultimate operating rate of our phase modulation
scheme was tested by running FG2 in Frequency-Shift Keying mode: its
frequency hopped between two values (corresponds to 0 or $\pi$ phase
delay) at 100 KHz rate. Note in this case, the equivalent
phase-encoding rate was 200 KHz. Here, strong, cw laser was input to
the system, and interference signal from Ch1 was detected with a
photo detector. Fig.5 shows the experimental result, where the sharp
rising edge ($\sim$500ns) indicates a potential phase modulation
rate of a few MHz could be achievable.

\begin{figure}[hbt]
\centering
\includegraphics[width=9cm]{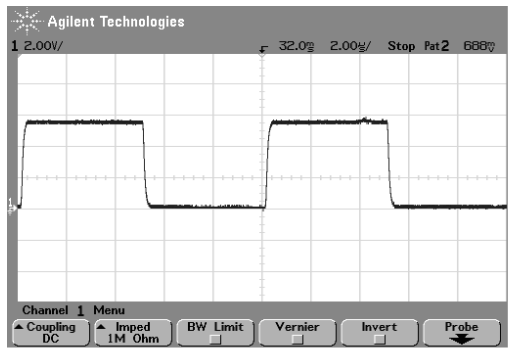}\\
\caption{The experimental results at 200 KHz Phase modulation rate.
The sharp rising edge ($\sim$500ns) indicates a potential phase
modulation rate of a few MHz could be achievable.} \label {result}
\end{figure}

\section{Discussion and Conclusion}

In a bidirectional QKD system, such as the ``plug \& play" system as
well as the system we propose here, Alice allows signals to go in
and out of her device, this opens a potential backdoor for Eve to
launch various Trojan horse attacks. In our simplified setup, where
only one AOM is used in Alice's side as a phase modulator, the
frequency of the laser pulse output from Alice depends on the
encoded phase information (due to the frequency shift induced by the
AOM). This may leak additional information to Eve and a naive
application of the unconditionally secure proof of BB84 to our
current experimental set-up is invalid.

In the standard BB84 protocol, Alice encodes phase information
\{$0$, $\pi/2$, $\pi$, $3\pi/2$\} by modulating the AOM at frequency
\{$f_0$, $f_0+\Delta f$, $f_0+2\Delta f$, $f_0+3\Delta f$\}. From
(2), the encoded phase depends on both the AOM's driving frequency
$f$ and the fiber length difference $\Delta L$. In principle, Eve
can build up a device with different fiber length, breaks into the
communication channel, and plays the ``intercept and resend" attack.
Suppose Eve uses her device to send laser pulses to Alice. Unaware
that the pulses are from Eve, Alice will still shift the frequency
of light by one of the values \{$f_0$, $f_0+\Delta f$, $f_0+2\Delta
f$, $f_0+3\Delta f$\}. By choosing a suitable fiber length
difference, Eve can re-mapping the encoded phase from \{$0$,
$\pi/2$, $\pi$, $3\pi/2$\} to \{$0$,$\delta$, $2\delta$,
$3\delta$\}, where $\delta$ is under Eve's control. In [24], we
describe detail of this ``phase-remapping attack" and proved that
Eve can learn the full information of the final key at the cost of
introducing a 14.6\% QBER. Note that this number is substantially
lower than the proved secure bound of 18.9\% for the standard BB84
protocol [25]. We remark that this loophole can be closed by
employing the phase modulator shown in Fig.1b, which introduces no
frequency shift (as the setup in Fig.2).

In conclusion, we propose a polarization-insensitive phase
modulation scheme and a stable Sagnac QKD system employing this
technique. Compared with previous Sagnac QKD schemes based on
polarization-sensitive phase modulators, our proposed system
demonstrated better performance over longer fiber. Preliminary
experimental results showed an interference visibility of 96\%
(99\%) for a 40km (10km) fiber loop at single photon level. With
this novel polarization-insensitive phase modulation scheme, we
expect the performance of many practical QKD systems can be greatly
improved.

\section*{Acknowledgment}



Financial supports from NSERC, CRC Program, CFI, OIT, PREA, and CIPI
are gratefully acknowledged.

\end{document}